\documentclass[twocolumn, longbib, numbers]{aastex701}
\usepackage{amsmath}
\usepackage{mathtools}

\begin{document}

\title{PSR J0614-3329: A NICER case for Strange Quark Stars}

\correspondingauthor{Swarnim Shirke}
\author[orcid=0000-0001-8604-5362]{Swarnim Shirke}
\affiliation{Inter-University Centre for Astronomy and Astrophysics, \\Post Bag 4, Ganeshkhind, Pune University Campus, Pune - 411007, India}
\email[show]{swarnim@iucaa.in}

\author[orcid=0009-0008-5947-3060]{Rajesh Maiti}
\affiliation{Inter-University Centre for Astronomy and Astrophysics, \\Post Bag 4, Ganeshkhind, Pune University Campus, Pune - 411007, India}
\email{rajesh.maiti@iucaa.in}

\author[orcid=0000-0002-0995-2329]{Debarati Chatterjee}
\affiliation{Inter-University Centre for Astronomy and Astrophysics, \\Post Bag 4, Ganeshkhind, Pune University Campus, Pune - 411007, India}
\email{debarati@iucaa.in}

\begin{abstract}
Precise measurements of neutron star masses and radii by the NICER mission impose important constraints on the nuclear equation of state. The most recent NICER measurement of PSR J0614-3329 reported an equatorial radius of $R_{eq} = 10.29^{+1.01}_{-0.86}$ km for a mass of $M = 1.44^{+0.06}_{-0.07} M_{\odot}$. Considering all the NICER measurements to date, we find substantial evidence using Bayesian hypothesis ranking for strange quark stars over physically motivated models of neutron stars compatible with this low radius. This provides a strong case for quark matter in neutron stars and also for the possible existence of strange quark stars, a consequence of the Bodmer-Witten hypothesis, suggesting that they could be considered among the population of compact stars during analyses of astrophysical data. Using a wide sample of equations of state, we report the nucleonic equations of state that best fit current observations and rule out one model of strange quark matter.
\end{abstract}


\section{Introduction} 

Neutron Stars (NSs), one of the most compact stellar objects in the Universe, sustain extreme physical conditions, with the density in their interior exceeding several times the nuclear saturation density ($n_{sat}$)~\citep{Lattimer2004}. The lack of first-principle calculations from the theory of strong nuclear force- quantum chromodynamics (QCD) to describe cold, dense nuclear matter limits our current understanding of the matter in the NS interior, making the composition of NSs an open question. Conjectured possibilities include the appearance of heavy baryons, hyperons, and kaon condensate in the NS core. 

Although the transition density is not well-known, another such possibility of a phase transition from nuclear matter to quark matter (QM) consisting of deconfined up, down, and strange quarks in the NS core (for details of QM in NS, see~\citet{Baym2018} and references therein), as predicted by QCD~\citep{Shuryak1980}, remains open. Such a quark core can be formed at the time of the formation of NSs triggered by supernovae~\citep{Sagert2009}, or a rise in central density during NS evolution~\citep{Glendenning1997} due to accretion of matter, spin-down, density fluctuations, or as a result of a merger. Recently, strong evidence for deconfined QM in NSs has been established~\citep{Annala2020, Annala2022,Annala2023}

It has been conjectured that such a hadron-quark phase transition can transform the entire NS into a self-bound compact strange quark star (SQS), a star composed entirely of strange quark matter (SQM)~\citep{Haensel1986,Alcock1986}. First studied by~\citet{itoh1970}, SQSs are a direct consequence of the SQM hypothesis~\citep{bodmer1971collapsed, witten1984cosmic} that says the 3-flavor strange quark phase could be a more stable form of matter compared to hadronic matter if energy per nucleon ($E/A$) of the SQM is lower than that of $^{56}$Fe, making SQM the true ground state of matter. SQSs can also be formed via primordial fluctuations or cosmological phase transitions in the early universe~\citep{witten1984cosmic}. 

\begin{figure*}
    \centering    \includegraphics[width=0.9\linewidth]{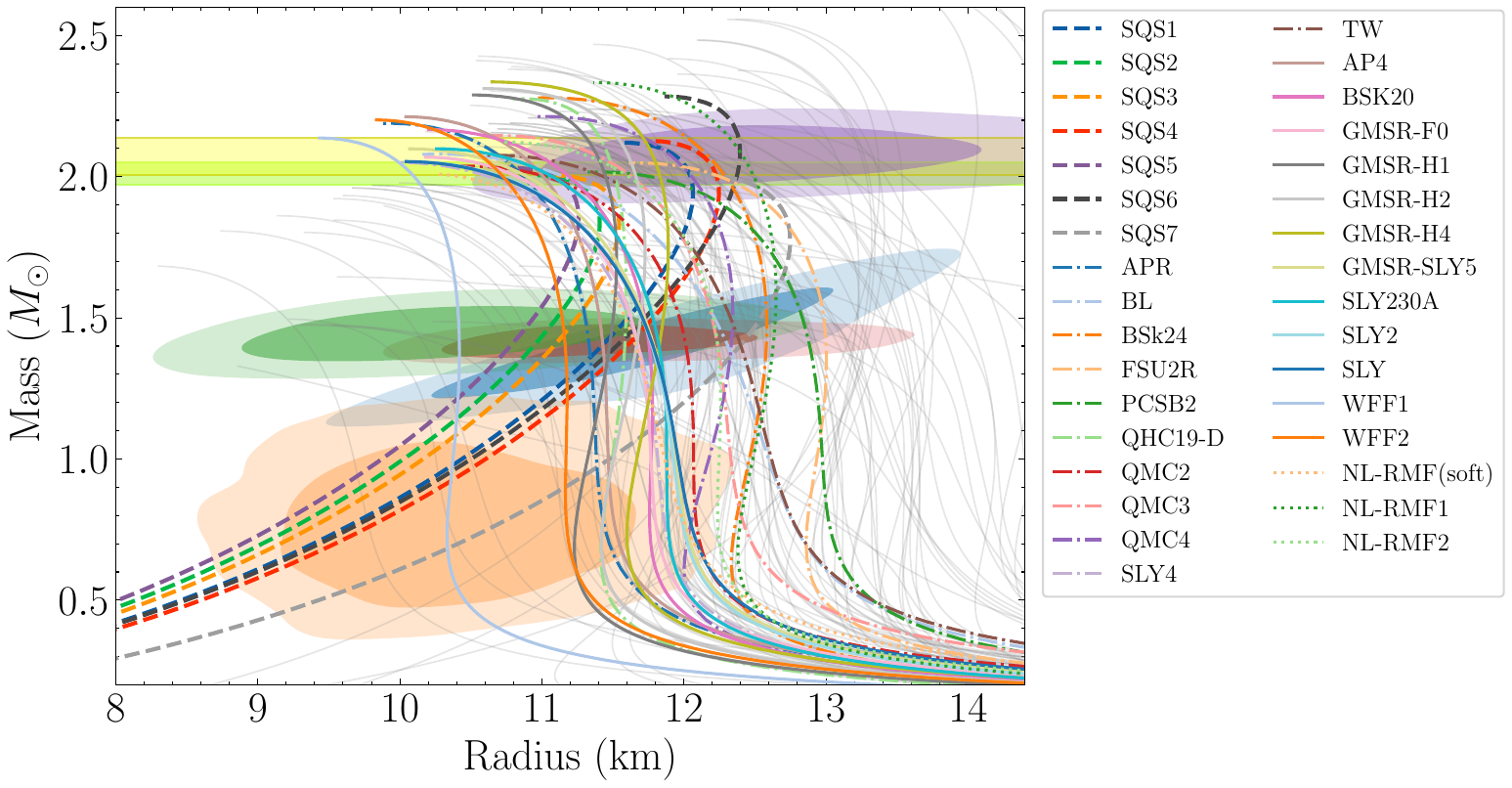}
    \caption{The mass-radius curves are displayed for 28 NS (dash-dotted, solid, and dotted lines) and 7 SQS (dashed line) EoSs used in this work (see text for details). The thin silver lines are all the EoSs in the LALSuite set. Various 68$\%$ (dark) and 95$\%$ (faint) mass-radius filled contours are overlaid: the orange patch corresponds to the HESS J1731-347~\citep{Doroshenko2022} observation, the purple patch to PSR J0740+6620~\citep{Salmi2024}, the red and blue patches to PSR J0437-4715~\citep{Choudhury2024}, and J0030+0451~\citep{Vinciguerra2024}, respectively, while the green patch corresponds to the latest observation of PSR J0614-3329~\citep{Mauviard2025}. The yellow and green bands correspond to mass measurements of the two heaviest pulsars known, $M=2.08^{+0.07}_{-0.07}$ of PSR J0740+6620~\citep{Fonseca2021} and  $M=2.01^{+0.04}_{-0.04}$ of PSR J0348+0432~\citep{Antoniadis2013}, shown only for comparison.}
    \label{fig:mr}
\end{figure*}

Detecting strangelets (SQM nuggets), SQSs, or phase transitions in NSs would have far-reaching physical implications on elementary particle physics as well as astrophysics. It holds the potential to delineate the phase transition density, the type of phase transition, constrain microphysics of cold QCD interactions, and reveal the true ground state of matter, establishing the existence of an entirely new class of stars. There have been extensive searches for signatures of SQM~\citep{Alford2019}, but the evidence for the existence of SQSs for all the candidates so far has been inconclusive~\citep{Weber2005}, keeping the question of their existence open. Observations of millisecond pulsars stable against the unstable $r$-mode oscillation have been one of the most clear indications of SQSs so far, as minimal NS models with nuclear composition fail to explain the expected damping~\citep{Madsen1998,Madsen2000,Alford2014}. Another smoking gun to settle the hypothesis is the simultaneous mass-radius measurement, as SQSs differ significantly from their NS counterparts (see Fig.~\ref{fig:mr}), forming configurations of high compactness and can sustain lower radii (especially at lower masses). 

The NS radius measurements have so far been scanty, uncertain, and involved model uncertainties. Simultaneous mass-radius measurements by the new Neutron star Interior Composition Explorer (NICER) instrument~\citep{Gendreau2016NICER} have been a milestone, providing radius estimates with an accuracy within $10\%$ from X-ray pulse profiles (PSR J0740~\citep{Salmi2024}, PSR J0030~\citep{Vinciguerra2024}, and PSR J0437~\citep{Choudhury2024}).
Observation of the gravitational waves (GWs) from the binary NS merger GW170817 provided an alternate measurement of tidal deformability ($\Lambda$), which can independently constrain the equation of state (EoS)~\citep{Abbott2017, Abbott2019}; however, since the current uncertainty is large, the promise of more precise measurements for NSs lies with the next-generation GW detectors. Although all these measurements, barring PSR J0740, ruled out stiffer EoSs, these have been compatible with a wide range of physically motivated NS models so far. 

The recent observation of the supernova remnant HESS J1731-347~\citep{Doroshenko2022} sparked speculations about the existence of SQSs owing to this low mass and radius~\citep{Horvath2023, Sagun2023, DiClemente2024} that SQSs naturally predict (see Fig.~\ref{fig:mr}). Although this measurement relies on the assumed atmosphere model and is sensitive to errors in distance and inferred age, the observation of this sub-solar mass supernova remnant still remains a possible hint in favor of SQSs.

The most recent result by NICER reported an equatorial radius of $R_{eq} = 10.29^{+1.01}_{-0.86}$ km for a mass of $M = 1.44^{+0.06}_{-0.07} M_{\odot}$ (68$\%$ credible interval), for  PSR J0614-3329~\citep{Mauviard2025} for their best fit headline model (ST+PDT) out of all others and is considered to be robust implying a tension with the earlier measurements claiming a lower radius for $1.4 M_{\odot}$ star. 
The hotspot model PDT-U resulted in the highest radius estimate in the entire study with $R_{eq} = 10.91^{+1.13}_{-1.0}$ km. Thus, at a 68$\%$ credibility, the radius estimate is less than $11.30$ km and in the least optimistic scenario, $12.04$ km. 
It was shown that this new observation rules out stiff EoSs and shifts the estimate of the radius of 1.4$M_{\odot}$ NS ($R_{1.4M_{\odot}}$) by $\sim300$ m, as also independently found in other analyses~\citep{SunnyNg2025, Biswas2025}. The implications of this have also been explored for other exotic matter~\citep{Alvarez2025, Shahrbaf2024} with a hint towards a possible phase transition~\citep {Tang2025}. None of the works considered the possibility of SQSs, which, as we show in this work, is best supported by the NICER measurements. Moreover, these studies resorted to either parametric EoSs like piecewise polytrope or speed of sound construction or the non-parametric Gaussian process, which do not take into account the underlying microphysics. Although the lower radii observations can be explained using parametric EoSs, it is not always possible to explain these using physically motivated models, making it important to validate the NICER results against available microscopic and phenomenological EoS models in light of the new measurement.

\section{Data and Models}\label{sec:analysis}
To demonstrate this, we make use of all the robust updated NICER simultaneous mass-radius measurements available, namely PSR J0740~\citep{Salmi2024}\footnote{https://doi.org/10.5281/zenodo.10519473} (henceforth J0740), PSR J0030~\citep{Vinciguerra2024}\footnote{https://doi.org/10.5281/zenodo.8239000} (henceforth J0030), PSR J0437~\citep{Choudhury2024}\footnote{https://doi.org/10.5281/zenodo.10886504} (henceforth J0437) and the latest measurement PSR J0614~\citep{Mauviard2025}\footnote{https://doi.org/10.5281/zenodo.15603406} (henceforth J0614). 
It has been shown for J0614 that ST+PDT (headline model) is strongly favored over the alternate PDT-U using multi-messenger EoS inference~\citep{Biswas2025}. 
Hence, we resort to the robust best-fit headline ST+PDT model for J0614 for the analysis here. 
The same was shown to be the case for PSR J0030~\citep{Luo2024}. Hence, we also use the ST+PDT result for J0030, which also ensures uniformity of analyses used.
The measurement of PSR J1231‑1411~\citep{Salmi2024ApJComplexCase} is known to be not robust and is thus not considered. $\Lambda$ measurement for GW170817~\citep{Abbott2017,Abbott2019} has a large uncertainty and provides only an indirect constraint for mass-radius through $C-$Love equations, which are distinct for NSs and SQSs~\citep{Yagi2013}. There are other less accurate measurements that can be taken into account~\citep{nattila2016, nattila2017} and other tests of QM~\citep{Alford2014}; but here, we focus on the implications of direct mass-radius observations from NICER X-ray pulse profiles, also maintaining uniformity of measurements. We have still considered only those SQS models with $\Lambda_{1.4M_{\odot}}<700$ (except SQS7, to be explained later) consistent with~\citet{Abbott2017,Abbott2019}. We consider the low mass HESS J1731-347 result~\citep{Doroshenko2022}\footnote{https://doi.org/10.5281/zenodo.6702216} (henceforth HESS) only to check the evidence it adds since it is already in favor of SQSs~\citep{Horvath2023, Sagun2023, DiClemente2024}; however, since the measurement is model dependent and is also incompatible with other XMM-Newton data~\citep{AlfordHalpern2023}, we restrict to the case without HESS to derive final conclusions. We respect the radio mass observations by considering only those EoSs that produce TOV maximum mass above the highest observed mass~\citep{Fonseca2021}. It is worth noting that this mass information is already encoded in the J0740 posterior. Thus, we have considered all the robust measurements providing direct mass-radius constraints, with the goal of studying the implications of the new NICER observation on realistic EoSs. 

We obtain a large sample of realistic EoSs from publicly available EoS repositories. We consider the 10 best possible and widely used cold nucleonic NS EoSs and one involving hadron-quark phase transition to investigate one case of a hybrid star from the CompOSE EoS database\footnote{https://compose.obspm.fr/} (dash-dotted curves in Fig.\ref{fig:mr}). We consider 3 EoSs from the phenomenological relativistic mean field theory that is widely used to model nuclear matter in NSs, with parameters optimized within their allowed uncertainties to ensure the best agreement
with astrophysical data (see Appendix~\ref{sec:ns_models}).
We also include all the EoSs from the LALSuite/LALSimulation\footnote{https://lscsoft.docs.ligo.org/lalsuite/lalsimulation/} (also used in BILBY) library, which is an exhaustive set used in the GW community for drawing inferences from GW events (thin silver lines in Fig.~\ref{fig:mr}). We draw 12 best candidates in light of the new low radius measurement, shown in solid lines in Fig.~\ref{fig:mr}. In our sample, we have considered the NS EoSs with low radii agreeing best with all the mass-radius contours, such that our choice is biased against our hypothesis to obtain a robust result (see Appendix~\ref {sec:ns_models} for details on the choice of the set of EoS).

For SQSs, we consider 4 EoSs from the widely used MIT Bag model~\citep{Farhi_Jaffe1984} (SQS1-4), 2 from the vector-enhanced vMIT Bag model~\citep{Klahn2015} (SQS5-6), and one from a model where the effect of running of the parameters $m_s$ and $\alpha_s$ given by~\cite{Fraga2005} (SQS7). The set SQS1-7 is not the best possible SQS EoSs and are such that they marginally satisfy $1\sigma$ $M-R$ contours (See Appendix~\ref{sec:sqs_models} for parameter choice); thus, the results we obtain can be considered not to be the most optimistic.

Fig~\ref{fig:mr} displays the $M-R$ curves for all the considered EoSs. We have also included the $68\%$ and $95\%$ filled contours for all four NICER observations, along with those from HESS. We observe that only GMSR-H1, H2, AP4, QHC19-D NS EoSs satisfy all the constraints within $1\sigma$, in contrast to SQS, which allows a wide range of EoSs consistent with observations, including HESS. Although they might still agree with $2\sigma$ contours, from this, we hypothesize that \textit{the NICER observations along with the new J0614 favor SQS models over NSs}.

\section{Analysis}\label{sec:bayesian_analysis}

To quantitatively test this hypothesis, we follow the Bayesian method of \textit{Hypothesis Ranking} as introduced in~\cite{Pozzo2013} for testing a discrete number of models (here EoSs) against each other. 
The odds ratio ($\mathcal{O}_{j}^i$) of EoS$_i$ against EoS$_j$ is given by
\begin{equation}
\mathcal{O}_{j}^i = \frac{P(\mathrm{EoS}_i | \vec{d})}{P(\mathrm{EoS}_j | \vec{d})},
\end{equation}
where $\vec{d}=(d_1,d_2,...,d_N)$ contains the list of all data to be considered. Using Bayes' theorem, this can be written as 
\begin{equation}
\mathcal{O}_{j}^i = \underbrace{\prod_{k=1}^N 
\frac{P(d_k|\mathrm{EoS}_i)}{P(d_k|\mathrm{EoS}_j)}}_{\text{Bayes factor}}
\times
\frac{P(\mathrm{EoS}_i)}{P(\mathrm{EoS}_j)},
\end{equation}
$P(\mathrm{EoS_i})$ is the prior on the EoS$_i$. Here, we assume all EoSs are equally likely with no prior bias, and the ratio of priors is 1. The problem of evaluating odds reduces to finding the Bayes factor
\begin{equation}
    B^i_j = \frac{\prod_{k=1}^NP(d_k|\text{EoS}_i)}{\prod_{k=1}^NP(d_k|\text{EoS}_j)}~.
\end{equation}
Each $d_k$ is the $k^{\text{th}}$ data set. We consider the earlier 3 NICEr results ($N=3$) in case A, all four NICER results ($N=4$) in case A, and the data for HESS ($N=5$) in case C for the analysis. 

For a given observation $d_k$ we compute the likelihood of $\text{EoS}_i$ ($P(d_k|\mathrm{EoS}_i)$) as 
\begin{equation}\label{eqn:likelihood}
P(d_k\,|\,\text{EoS}_i) = 
\int_M \int_R dM\,dR\, P(d_k\,|\,M,\,R)\,P(M,R\,|\,\text{EoS}_i) \;.
\end{equation}
To compute $P(d_k\,|\,M,\,R)$, we model it using a Gaussian Kernel Density Estimator (KDE) to obtain a smooth probability density distribution over the mass–radius plane from the posterior reported by each observation. Since an EoS gives a curve in $M-R$, we have $P(M,R\,|\,\text{EoS}_i) = P(M\,|\, \text{EoS}_i)\delta(R-R(M\,|\,\text{EoS}_i)$. As follows in~\cite{Biswas2022}, given the small number of observations, we consider a uniform prior for 
\begin{equation}
P(M\,|\,\mathrm{EoS}_i) = 
\begin{cases}
\dfrac{1}{M_{max} - M_{min}} ~~~ \mathrm{if~} ~M_{min} \leq M \leq M_{max} \\
~~~~~~~~~~~~0  ~~~~~~~~~~~~~~\text{otherwise}
\end{cases}
\end{equation}

Here, $M_{max}$ is the maximum TOV mass of the EoS. We used $M_{min}=0.5M_{\odot}$, keeping in mind the HESS observation. Since the $1\sigma$ HESS contour lies $\gtrsim0.5M_{\odot}$, this ensures that all the posteriors are well sampled. Using this Eq.~\ref{eqn:likelihood} becomes
\begin{equation}
P(d_k\,|\,\mathrm{EoS}_i) = 
\frac{
    \int_{M_\mathrm{min}}^{M_\mathrm{max}} dM\, P\big(d_k \,|\, M, R(M\,|\,\mathrm{EoS}_i)\big)
}{
    M_\mathrm{max} - M_\mathrm{min}
}.
\end{equation}

We compute $\prod_{k=1}^nP(d_k|\text{EoS})$ for each model for all the cases.. A positive value of $\log_{10}{B}$ (or $B>1$) means that SQS is favored over the NS model. The higher a Bayes factor, the higher the preference. The factor is commonly interpreted as follows~\citep{jeffreys1939theory}: (i) $\log_{10}{B} \ge 2$ means decisive evidence in favor of SQS (ii) $1 \le \log_{10}{B} <2$
 means strong evidence (iii) $1/2 < \log_{10}{B} < 1$ means substantial evidence and (iv) $\log_{10}{B}\le 1/2$ means insubstantial evidence. There are some alternative scales used for interpretation given by~\citet{Kass1995} (used in GW analyses) and~\citet{lee2014bayesian}, but the interpretation of our conclusions remains unaltered in all these scales as well.

We compute the Odds ratio of all 7 SQS EoSs to all 26 nucleonic EoSs to obtain $26\times 7$ Bayes factors for SQS EoS over a NS EoS (see tables in Appendix~\ref{sec:extended_data}) for three cases: Case A) Only the earlier NICER results (J0030+J0740+J0437); Case B) including the new NICER measurement (J0614) and Case C) further including HESS (for comparison). As could be guessed from Fig.~\ref{fig:mr}, we find that SQS7 is not a good fit to the observations (see Table.~\ref{tab:full_nicer_only}); hence, we rule it out and exclude it from further analysis. We conclude that the MIT and vMIT models provide a better fit to the data. 

For the considered sample, we can further calculate the probability that the data can be explained by SQS as a separate check for our hypothesis. Following~\citet{Kass1995},  
for the given set of discrete models, we can write using Bayes' theorem

\begin{align}
    P(SQS|\vec{d}) &= \frac{P(\vec{d}|SQS)P(SQS)}{P(\vec{d})} \\ 
    &= \frac{P(\vec{d}|SQS)P(SQS)}{P(\vec{d}|SQS)P(SQS) + P(\vec{d}|NS)P(NS)} 
\end{align}
Treating both models to be equally likely to derive the implications of the NICER data alone, we get
\begin{align}
    P(SQS|\vec{d}) &= \frac{P(\vec{d}|SQS)}{P(\vec{d}|SQS) + P(\vec{d}|NS)} \nonumber \\
    &= \frac{\sum_i P(\vec{d}|EoS_i,SQS)P(EoS_i|SQS)}{\left( \splitfrac{\sum_i P(\vec{d}|EoS_i,SQS)P(EoS_i|SQS)}{+\sum_i P(\vec{d}|EoS_i,NS)P(EoS_i|NS)}\right)}
\end{align}
where the summation is over all EoSs for the discrete case considered here. We consider all EoSs, under the assumption of a model, to be equally likely. Hence, here, $P(EoS_i|SQS)$ is 1/6 for SQS EoSs and 0 for NS and $P(EoS_i|NS)=1/25$ for NS EoS and 0 for SQSs. These factors also take into account the effect of sample size. Since we have considered only those NS EoSs for which $P(\vec{d}|EoS_i,NS)$ is large, adding more NS EoSs is likely to raise $P(SQS|\vec{d})$. It can be checked that  $P(NS|\vec{d}) = 1 -P(SQS|\vec{d})$. The corresponding Bayes factor equivalent is then given by $P(SQS|\vec{d})/P(NS|\vec{d}) = P(SQS|\vec{d})/(1-P(SQS|\vec{d}))$.

\section{Results}\label{sec:results}

\begin{figure*}
    \centering    \includegraphics[width=\linewidth]{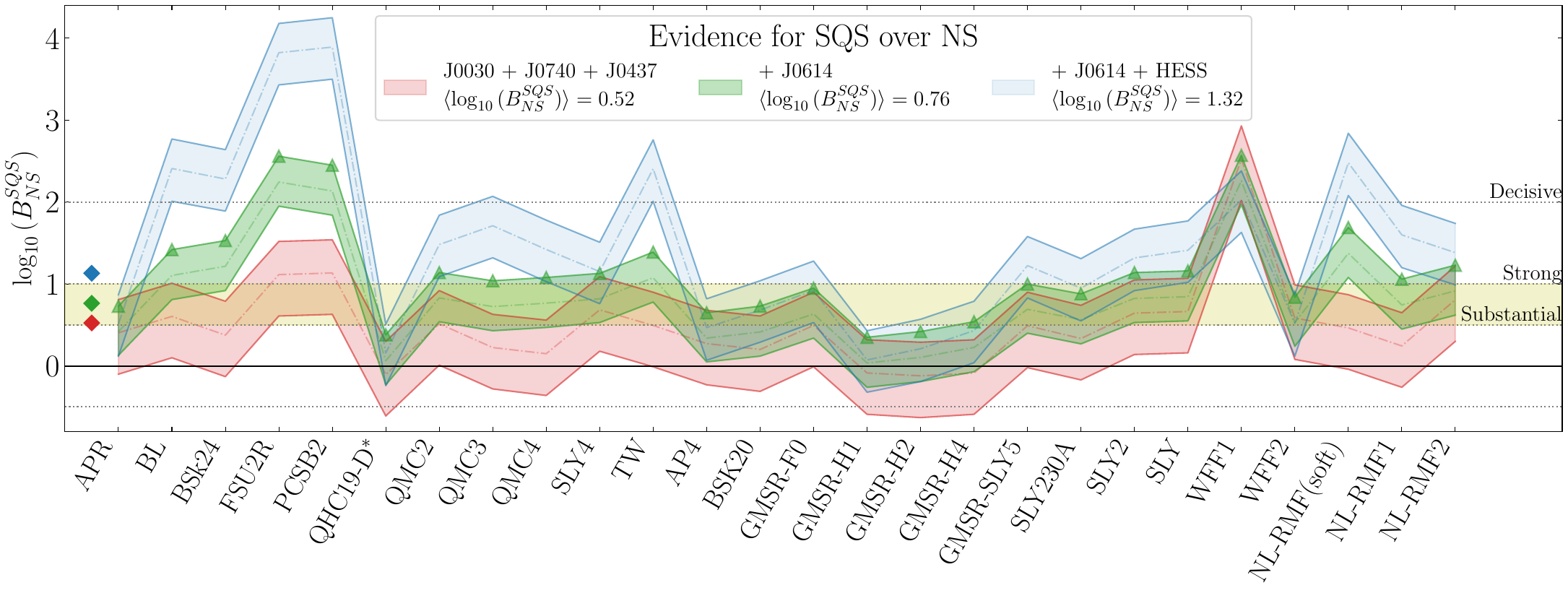}
    \caption{The bands of Bayes factor $\log_{10}{(B^{SQS}_{NS}})$ formed by SQS1-6 against each NS model EoS. The red, green, and blue bands correspond to Case A, B, and C, respectively (see text). The upper, middle, and lower curves in each band mark the highest, mean, and the lowest Bayes factor from SQS1-6. The curves with green triangles can be considered as lower bounds for each case. The diamonds on the left are the median values for each band considered as the benchmark in this study. The factor is interpreted as follows~\citep{jeffreys1939theory}: (i) $\log_{10}{B} \ge 2$ means decisive evidence in favor of SQS (ii) $1 \le \log_{10}{B} <2$ means strong evidence (iii) $1/2 < \log_{10}{B} < 1$ means substantial evidence and (iv) $\log_{10}{B}\le 1/2$ means insubstantial evidence; the horizontal dotted lines mark these boundaries. The region where substantial evidence exists is highlighted in yellow. *QHC19-D is a hybrid EoS model and is not considered in the calculations.}
    \label{fig:evidence}
\end{figure*}

We plot the data from Appendix~\ref{sec:extended_data} for the three cases in Fig.~\ref{fig:evidence} with red, green, and blue bands, respectively. The lower and upper bounds of the band mark the Bayes factor for the lowest and the highest factors among SQS1-6, for each nuclear EoS. The most important of these is the top curve of the green band (marked with triangles) since it considers all the robust NICER measurements and still corresponds to factors of not-the-best SQS model with the best of the NS models, meaning that it provides an underestimate. We find that the majority of these points suggest substantial evidence for SQS EoS, if not strong. The exceptions are QHC19-D, GMSR-H1, and H2. The negative values in favor of NS models do not imply that they are favored over SQSs. It only means that in our analysis, we find these to be the best models for NSs. The dotted line within the band is the mean average value for each nuclear EoS. If we consider this as the benchmark, to be pessimistic, we find APR, AP4, BSK20, and GMSR-4 are the next best NS models. Out of these, QHC19-D was considered as a check for hybrid EoS, indicating that a model with a quark core is favoured as it performs better than most nucleonic models, and that hybrid stars remain a viable candidate. Since our focus is on SQSs and NSs, we do not consider it in further analysis. 

In all the other cases, the not-the-best SQS EoSs are favored over the best considered nucleonic EoS. We conclude that there exist multiple SQS models that the analysis prefers over all the good-fitting, physically motivated nucleonic EoSs so far. Since the mean of Bayes factor is not a meaningful statistic to draw inferences, we report the median Bayes factor ($\langle\log_{10}{(B^{SQS}_{NS}}\rangle$) for each band on the left side of the plot in diamonds (also in the legend) as a measure of evidence in majority of cases. The yellow band marks the region with substantial statistical evidence for SQS over NSs as per the standard scale~\citep{jeffreys1939theory}.

We find that even in case A) $\langle\log_{10}{(B^{SQS}_{NS}}\rangle=0.52$, meaning that even before the J0614 measurement, there was substantial evidence for SQS. The reasons for this could be: i) J0437 already disfavors stiff nuclear EoSs at $1.4M_{\odot}$, and it is the most precise and constraining measurement of all the NICER results, being the nearest pulsar~\citep{Johnston1993} and brightest source in the NICER band~\citep{Choudhury2024}; ii)The shape of J0030 contour due to degeneracy in $M$ and $R$ is aligned roughly along constant compactness ($M/R$) axis, as followed approximately by SQSs; iii) As compared to other measurements, the heavy pulsar J0740 admits a larger radius at $2M_{\odot}$. This rise in the radius with mass is consistent with the nature of $M-R$ curves of self-bound SQSs. Although this factor is too close to the boundary of the two regions to draw a confident conclusion, we find that there already was marginal evidence for QM in compact stars from NICER.

Including the new result J0614 (case B) clearly makes the evidence substantial ($\langle\log_{10}{(B^{SQS}_{NS}}\rangle=0.76$) in favour of SQSs. This is the other main finding of this work, supporting the hypothesis that the new NICER measurement J0614 has gathered substantial evidence in favor of SQS models over NSs. In case C), where we included HESS, we observe that all the factors (except for WFF1-2) are boosted, making the case for SQSs strong ($\langle\log_{10}{(B^{SQS}_{NS}}\rangle=1.32$). The conclusion remains unaltered if we consider a different scale given by~\citet{Kass1995} for interpreting Bayes factors as is used in the GW community. Note, we have made use of median values considering all the SQS1-6 to have a conservative underestimate; if we only consider the upper boundary of the green band, which is also an underestimate, the evidence is even stronger.

\begin{figure}
    \centering    \includegraphics[width=\linewidth]{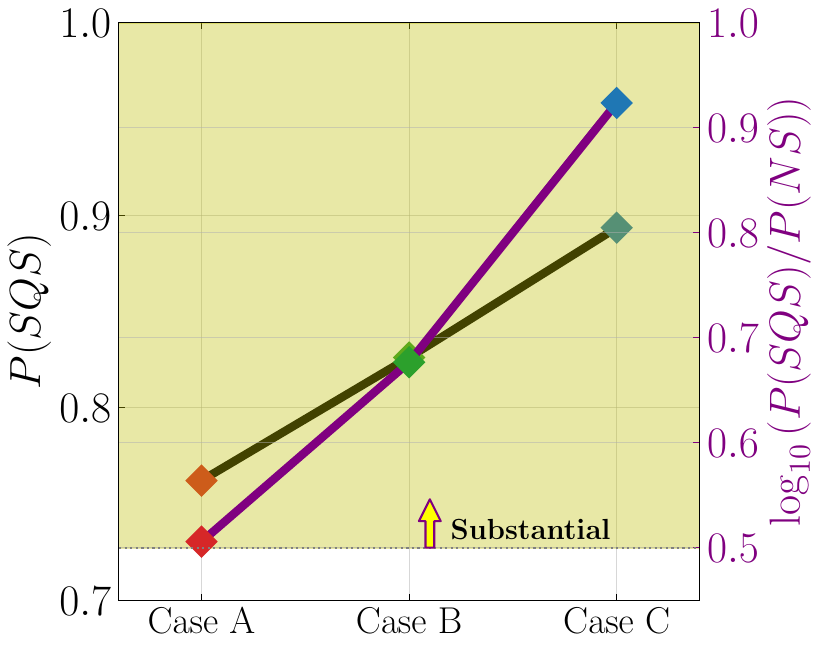}
    \caption{The probability of SQSs (see Sec.~\ref{sec:bayesian_analysis} for the calculation) to be the correct model given the NICER data and given the sample is shown by markers for each case connected by a black line. The purple line with an alternate axis is the corresponding Bayes equivalent. The yellow region highlights the region that can be considered a substantial evidence $1/2 < \log_{10}B <1$.}
    \label{fig:prob_evidence}
\end{figure}

As an alternative measure for evidence, we plot the total probability $P(SQS|\vec{d})$ and corresponding Bayes equivalent $\log_{10}{(P(SQS|\vec{d})/P(NS|\vec{d}))}$ (See Sec.~\ref{sec:bayesian_analysis}) for the entire sample of EoSs, instead of pairwise values, in Fig.~\ref{fig:prob_evidence} for each of the three cases. This depends on the chosen sample, but since we have chosen the optimistic cases for NSs, this is again an underestimate resulting in $\log_{10}{(P(SQS|\vec{d})/P(NS|\vec{d}))}=0.51,0.68,0.92$ for cases A, B and C, respectively. This provides an additional check for our conclusion that NICER observations provide substantial evidence for SQSs.

\section{Conclusions}\label{sec:conclusions}
To summarize, we have presented a strong case in support of the possible existence of SQSs. These have been considered as a possible alternate explanation of NSs for decades and have not been explored in light of state-of-the-art NICER observations. We found that NICER observations provide substantial evidence of SQSs over minimal NS models with $npe\mu$ matter, hinting towards the possibility of their existence. This evidence existed before only marginally, but the recent NICER analysis of PSR J0614-3329~\citep{Mauviard2025} with a low radius for NSs makes the conclusion confident. For completeness, we showed that the inclusion of HESS results makes the evidence strong. Since the explanation of the $r$-mode instability window of fast-spinning pulsars~\citep{Madsen1998, Madsen2000, Alford2014}, and theoretical evidence of QM in compact stars from perturbative QCD~\citep{Annala2020, Annala2022, Annala2023}, this is the next strong hint suggesting SQSs or QM in NSs. We reserve that confirmation from other methods is essential for conclusively establishing the existence of SQS; however, the currently available observational data still allows for the presence of QM in compact stars~\citep{Alford2007, Alford2019}. 

Unlike the earlier analyses, which resorted to parametric EoSs that can be fitted to low radii solutions, we showed that it is not necessarily possible to do so considering physically motivated NS EoSs. This does not mean that such NS EoSs are physically forbidden, but that NS EoSs that have been derived so far from microphysics and used in astrophysical analyses fall short in explaining the observations. On the other hand, we show that a range of phenomenological SQS models are consistent with all the NICER measurements, with Bayesian analysis providing substantial evidence in favor of SQSs when physically motivated models for both are considered. 

Albeit considering discrete cases, we arrive at this conclusion even after considering the best cases for NS models with the not-the-best SQS models, meaning that the evidence reported in this work is an underestimate. A full Bayesian model selection analysis can be performed assuming particular models for SQS and NSs; however, such an analysis would suffer from model bias, and we only expect stronger evidence from such an analysis. On the contrary, we have considered a wide range of NS models in this analysis. Hence, we propose that there is enough reason for SQSs, as proposed by~\citet{Haensel1986,Alcock1986}, to be included in analyses of astrophysical data as an important alternative scenario for NSs. Assuming no systematic bias in the X-ray pulse profile analyses/hotspot models, this also highlights the need for a better understanding of the underlying physics of NSs, else considering pulsars to be purely nucleonic NSs might lead to an incorrect interpretation of GW and other multimessenger data from pulsars. 
One caveat is that the NICER analyses make use of the quasi-universal relationship between spin and oblateness~\citep{AlGendy2014} determined with nucleonic EoS only (private communication by Sebastien Guillot). This may lead to an inconsistency issue; however, we do not expect the results to be affected significantly, given the low uncertainties in the relation. Nevertheless, this only reinforces the argument that analyses should incorporate SQS models in the future, as proposed in this work.

Comparing the Bayes factor across different EoS models, we still found some good NS models (not conclusively better than SQSs) and found that GMSR-H1 and H2 are the NS models that best explain the NICER data, followed by APR, AP4, BSk20, and GMSR-H4. The case of QHC19-D highlights the possibility of the existence of hybrid stars. In particular, the cases of SQSs with crusts~\citep{Jaikumar2006}, an early strong first-order phase transition hybrid star~\citep{Gartlein2025}, a slow stable hybrid star~\citep{Mariani2024}, and twin stars~\citep{Benic2015} remain an interesting alternative. Nevertheless, if favoured, these are still cases of QM, strengthening the evidence for QM in compact stars. \cite{Mauviard2025} reported a bimodal distribution for an $R_{1.4M_{\odot}}$, which hints towards the possibility of twin stars, although it is too early for a conclusive analysis given the current NICER uncertainties.

On a fundamental level, although it may seem that the analysis hints towards stability of 3-flavor QM, defending the SQM hypothesis, we emphasize that more precise measurements, especially of low-mass and same-mass NSs, and better GW observations are essential to decisively establish the existence of SQSs and/or QM and to distinguish between various models of compact stars with QM~\citep{Most2018, Bauswein2019,Pradhan2024Twin}. Future NICER observations and next-generation multimessenger missions such as eXTP (2030, \cite{Watts2019,Li2025}), STROBE-X (2030, \cite{Ray2019}), NewAthena (2037, \cite{Cruise2025}), Einstein Telescope~\citep{Maggoire2020ET}, Cosmic Explorer~\citep{Reitze2019CE}, and NEMO~\citep{Ackley2020NEMO} hold great promise to distinguish between various models for compact objects. Finding a hint of QM in compact stars, as illustrated here, would be a significant addition to the list of achievements of NICER as a science mission, impressing the importance of these future missions. This work also paves the way for future investigations of additional degrees of freedom, such as hyperons, meson condensates, $\Delta-$baryons, H-dibaryons, dark matter, and more nuclear interactions, probing signatures of other exotic physics.

\begin{acknowledgments}
The authors thank Bikram Keshari Pradhan, Suprovo Ghosh, Anna Watts, and Sebastien Guillot for useful comments that improved the manuscript significantly.
\end{acknowledgments}




%
\facilities{\textit{NICER}}

\software{\texttt{CompOSE}\footnote{\label{note1}https://compose.obspm.fr/}, \texttt{LALSuite}/\texttt{LALSimulation}\footnote{https://lscsoft.docs.ligo.org/lalsuite/lalsimulation/}, \texttt{NumPy}~\citep{harris2020arrayNumpy}, \texttt{SciPy}~\citep{2020SciPy-NMeth}, \texttt{Matplotlib}~\citep{Hunter:2007Matplotlib}, \texttt{Seaborn}~\citep{Waskom2021Seaborn}, Scienceplots~\citep{SciencePlots} }

\appendix



\section{EoS Models}\label{sec:methods}
Here, we describe the EoS models used, along with the reasons for considering these.

\subsection{NS Models}\label{sec:ns_models}
CompOSE, an EoS library widely used in nuclear and NS studies, contains a wide range of microscopic and phenomenological EoSs motivated by physics. We have taken 10 nuclear cold NS EoSs containing normal $npe\mu$ matter that are widely used in literature:
APR4~\citep{APR1998}, SLy4~\citep{Chabanat1998}, BL~\citep{Bombaci2018}, BSk24~\citep{Goriely2013}, TW~\citep{Typel1999}, FSU2R~\citep{Negreiros2018}, PCSB2~\citep{Pradhanzeta2023}, QMC2, QMC3, and QMC4~\citep{Alford2022} and one EoS containing hadron-quark crossover QHC19-D~\citep{Baym2019}, 
obtained from CompOSE EoS repository.
The hybrid QHC EoS is considered to check one case of a hybrid star, which involves a transition from hadronic to quark phase transition via a crossover and is a good candidate since it contains a soft nucleonic EoS at low densities and the crossover stiffens the EoS, unlike other hybrid star models, due to strongly interacting quark matter and satisfies the mass constraint. These nucleonic EoSs can fit the observations and have not considered EoSs with $R_{1.4M_{\odot}}\gtrsim13$ km, such that our choice is biased against our hypothesis to obtain a robust result. 12 EoSs from LALSimulation are chosen such that $R_{1.4M_{\odot}}\lesssim12$, as CompOSE resulted in only a few low radii EoSs, making sure the models are not repeated. These are: BSk20, GMSR-F0, GMSR-H1, GMSR-H2, GMSR-H4, GMSR-SLY5, SLY230A, SLY2, SLY, WFF1, and WFF2. We also use EoSs from the widely-used phenomenological non-linear relativistic mean-field (NL-RMF) model~\citep{Chen2014,hornick2018} where the interactions between the nucleons are mediated via the exchange of mesons. Variation in the interaction couplings allows for a wide range of EoS that agree with the empirical nuclear saturation parameters. We have optimized the EoS parameters within their allowed uncertainties to ensure the best agreement with the observations. For this, we consider the softest possible EoS (NL-RMF(soft)) consistent with the low-energy chiral effective field theory calculation~\citep{Drischler2016} and generate a two-solar-mass star as observed from radio measurements from the posterior generated in~\cite{Ghosh2022EPJA}. The EoSs on the stiffer part are expected to be ruled as per all the analyses~\citep{Mauviard2025,SunnyNg2025}. Hence, apart from this, we consider two other parameterizations (NL-RMF1 and NL-RMF2) that agree well with all the parameters, passing through $1\sigma$ contours of all the considered constraints except HESS and J0614, as no EoSs for this model agree with all these. For both NL-RMF1 and NL-RMF2, we fixed the saturation density ($n_{\text{sat}} = 0.15\,\text{fm}^{-3}$), binding energy per nucleon ($E/A = -16\,\text{MeV}$), isoscalar incompressibility ($K_{\text{sat}} = 240\,\text{MeV}$), and symmetry energy ($J = 31\,\text{MeV}$) at their median values within the range of uncertainties. Whereas the slope of the symmetry energy parameter ($L = 40\,\text{MeV}$), which correlates with $R_{1.4\,M_\odot}$~\citep{Ghosh2022EPJA}, is kept low, and the Dirac effective mass ($m^*/m$), which anti-correlates with the maximum mass, is set to $m^*/m = 0.70$ for NL-RMF1 and $0.75$ for NL-RMF2, within their current uncertainties, to produce a smaller radius while still satisfying the high-mass constraint from J0740. Thus, we consider a total of 26 nucleonic EoSs.


%


\subsection{SQS Models}\label{sec:sqs_models}
For SQSs, we consider two widely used phenomenological models: The MIT Bag model~\citep{Farhi_Jaffe1984} (SQS1-4) and the vector-enhanced vMIT Bag model~\citep{Klahn2015, Lopes2021a} (SQS5-6).  For each, we consider EoSs with the softest configuration that satisfy the two-solar-mass limit and a stiffest configuration agreeing with $1\sigma$ contours of all the considered constraints. Intermediate cases are also possible, but we consider the two extreme cases so as to get the results on the extreme ends, as we expect the intermediate ones to be even strongly favored.  The EoSs stiffer than these are already expected to be ruled out. Apart from this, one model where the effect of running of the parameters $m_s$ and $\alpha_s$ given by~\cite{Fraga2005} is also considered (SQS7). The models SQM1-3 by~\cite{Lattimer2001} have been used in literature, but we do not consider them separately as they are examples of the bag model itself and do not reach the two-solar-mass limit. Thus, we have a total of 7 SQS EoSs. 
For all the quark models, we make sure that the energy requirements of the SQM hypothesis, given by,
\begin{equation}
    \frac{E}{A}\bigg|_{N_f=3} < \frac{E}{A}\bigg|_{^{56}\text{Fe}}=930\text{ MeV} < \frac{E}{A}\bigg|_{N_f =2}~,
\end{equation}
is satisfied as only then are SQSs physically viable. The details of these SQS models are given below.

\subsubsection{MIT Bag model}
The EoS that was used to model SQS1-4 is the widely used MIT Bag model~\citep{Farhi_Jaffe1984, Glendenning_book1996} for ungapped 3-flavor QM along with a first-order correction in the strong interaction coupling and finite strange quark mass $m_s$. The full EoS as proposed by~\citet{Farhi_Jaffe1984} reduces to
\begin{align}
P &= \frac{a_4}{4\pi^2}\left( \mu_d^4 + \mu_u^4 + \mu_s^4 \right)
- \frac{3m_s^2 \mu_s^2}{4\pi^2} \\
&+ \frac{3m_s^4}{32\pi^2}\left( 3 + 4 \log \left( \frac{2\mu_s}{m_s} \right) \right)
- B + \frac{\mu_e^4}{12\pi^2},\nonumber
\end{align}
when expanded in powers of $m_s/\mu$~\citep{Alford2005, Alford2012}. Here, $B$ is the bag constant responsible for deconfinement and accounting for non-perturbative QCD background, $\mu_u$, $\mu_d$, and $\mu_s$ are the chemical potentials of the up, down, and strange quarks, and $\mu=(\mu_u+\mu_d+\mu_s)/3$ is the quark chemical potential. The energy density is calculated using the Gibbs-Duhem equation $\epsilon = \sum_i\mu_in_i-P$. If $m_s$ is considered to be zero ($m_s \ll \mu$) the EoS becomes independent of $a_4$ and takes a simpler form
\begin{align}
    P = (\epsilon-4B)/3~.
\end{align}
This EoS is consistent with the conformal speed of sound value of $c_s^2=dP/d\epsilon=1/3$. The effect of color-superconductivity leading to an energy gap can further be considered. 

For the MIT Bag model, we consider two cases: i) where the mass of the strange quark $m_s=0$ MeV. This results in the simplest form of the Bag model EoS $p=(\epsilon-4B)/3$. We mark SQS1 and SQS2 corresponding to EoSs with $B^{1/4}=140$ MeV and $B^{1/4}=145$ MeV, respectively; ii) where $m_s=100$ MeV. Here, the role of the strong-interaction parameter $a_4$ becomes important. We mark the two EoSs with parameters {$a_4=0.7$, $B^{1/4}=141$ MeV} and {$a_4=0.8$, $B^{1/4}=137$ MeV} by SQS3 and SQS4.

\subsubsection{vMIT Bag model}
As the original MIT Bag model does not contain the vector interaction, it struggles to produce massive stars, especially when hybrid stars are considered. A vector channel $V_\mu$, analogous to the $\omega$ meson in Quantum Hadrodynamics
(QHD) models~\citep{Chen2014,hornick2018}, is introduced in the vMIT Bag model to resolve this issue. We consider this model for SQS5 and SQS6. The model also introduces a quartic term for the vector field as a correction for the EoS at high density to mimic the Dirac sea contribution. The Lagrangian density is given by~\citep{Klahn2015,Gomes2019,Lopes2021a},\\

\begin{align}
    \mathcal{L} &= \sum_{u,d,s} \{\Bar{\psi}_q [\gamma^\mu (i\partial_\mu - g_{qqV_\mu}) - m_q]\psi_q - B\} \nonumber\\
    &+ \frac{1}{2}m_V^2V_\mu V^\mu +  \frac{b_4}{4}(g^2 V_\mu V^\mu)^2
\end{align}
where $\psi_q$ is the Dirac spinor for the quarks, $m_V$ and $m_q$s' are the masses of the vector field and the quarks, respectively. The interaction strength parameters are related as
\begin{align*}
    g_{uuV} &= g_{ddV} = g \\
    g_{ssV} &= X_v g_{uuV} = X_v g_{ddV}
\end{align*}
where $X_v$ is the coupling ratio of the strange quark and up/down quark to the vector field. The universal quark-quark interaction suggests $X_v = 1.0$, whereas the symmetry group SU(6) fixes this parameter as $X_v = 0.4$. This model effectively has three parameters: $B^{1/4},\;G_v = (\frac{g}{m_v})^2$, and $b_4$. The interaction strength parameter $b_4$ of the quartic coupling helps to stiffen or soften the EoS depending on its negative and positive values, respectively.

For the vMIT Bag model, depending on the vector interaction strength parameter $G_v$, we choose two sets as SQS5 ($G_v=0.1 \;\text{fm}^2$, $B^{1/4}=146$ MeV) and SQS6 ($G_v=0.2 \;\text{fm}^2$, $B^{1/4}=141$ MeV). For both the parameterizations, we consider the coupling ratio of strange quark and up/down quark to the vector field, $X_v=0.4$, as suggested by the symmetry group SU(6). The quartic vector self-interaction parameter $b_4$ is set to zero, as its inclusion makes the EoS excessively stiff at high densities, causing the speed of sound to become acausal, which is unphysical. 

\subsubsection{pQCD Model}

A model motivated by not just first-order corrections in strong interaction coupling $\alpha_s$ and finite $m_s$ but also renormalization group running of these quantities derived in~\citet{Fraga2005} is used for SQS7. The thermodynamic potential ($\Omega=\Omega^{(0)}+\Omega^{(1)}=-P$) contains the full first order correction ($\Omega^{(1)}$) over the free Fermi temp ($\Omega^{(0)}$)~\citep{Farhi_Jaffe1984, Glendenning_book1996, Hernandez2024}.
$\Omega$ depends explicitly on the renormalization scale ($\bar{\Lambda}$) and also implicitly via $\alpha_s(\bar{\Lambda})$ and $m_s(\bar{\Lambda})$ given by
\begin{align}
\alpha_s(\bar{\Lambda}) &= \frac{4\pi}{\beta_0 L} \left[ 1 - 2 \frac{\beta_1}{\beta_0^2} \frac{\ln L}{L} \right],\\
m_s(\bar{\Lambda}) &= \hat{m}_s \left( \frac{\alpha_s}{\pi} \right)^{4/9} \left[ 1 + 0.895062 \frac{\alpha_s}{\pi} \right]~.
\end{align}
Here, $L = 2\ln(\bar{\Lambda}/\Lambda_{\mathrm{MS}})$, $\beta_0 = 11 - 2N_f/3$, and $\beta_1 = 51 - 19N_f/3$ and $N_f = 3$ for SQSs. The scale $\Lambda_{\mathrm{MS}}$ and the invariant mass $\hat{m}_s$ are fixed according to experimental observations from Particle Data Group to give $\alpha_s \simeq 0.3$ and $m_s \simeq 100~\mathrm{MeV}$ at $\bar{\Lambda} = 2~\mathrm{GeV}$. This give $\Lambda_{\mathrm{MS}} \simeq 380~\mathrm{MeV}$ and $\hat{m}_s \simeq 262~\mathrm{MeV}$. We are only left with the renormalization scale $\bar{\Lambda}$, usually taken to be $2\mu$ or $3\mu$, which dictates the EoS. We consider the with $\bar{\Lambda}=2\mu$, $\mu$ being the quark chemical potential as done in~\cite{Fraga2001, Hernandez2024}. $\bar{\Lambda}=3\mu$ results in an even stiffer EoS and is hence not considered.



\section{Extended data}\label{sec:extended_data}

We report the Bayes factors for all the $26\times7$ combinations for all three cases in Table~\ref{tab:old_nicer_only}, Table~\ref{tab:full_nicer_only}, Table~\ref{tab:full_nicer_hess} in a matrix form. Each value in the matrix is $\log_{10}{B^i_j}$, where $i$ corresponds to the SQS EoS marked by the column name and $j$ is the nucleonic EoS marked by the row name. The last row gives the mean of the logarithm of Bayes factors for a SQS1-7 averaged over all nucleonic EoS. 


\begin{deluxetable*}{r|ccccccc}
\digitalasset
\tablewidth{100in}
\tablecaption{Log of Bayes factor ($\log_{10}{B}$) for all 7 SQS EoSs against all 26 NS EoSs using posteriors of ealier NICER measurements (J0740, J0030, J0437) only. The last row in the mean average of the logarithm of Bayes factors for a SQS1-7 over all nucleonic EoSs. $^*$QHC19-D is a hybrid star EoS. \label{tab:old_nicer_only}}
\tablehead{
\colhead{Model} & \colhead{SQS1} & \colhead{SQS2} & \colhead{SQS3} & \colhead{SQS4} & \colhead{SQS5} & \colhead{SQS6} & \colhead{SQS7}
}
\startdata
APR          &  0.81  & -0.00  &  0.16  &  0.78  &  -0.10  &  0.78  &  -0.39 \\ 
BL           &  1.01  &  0.20  &  0.36  &  0.98  &   0.10  &  0.98  &  -0.19 \\
BSk24        &  0.79  & -0.03  &  0.13  &  0.75  &  -0.13  &  0.75  &  -0.42 \\
FSU2R        &  1.52  &  0.71  &  0.87  &  1.49  &   0.61  &  1.49  &   0.32 \\
PCSB2        &  1.54  &  0.73  &  0.89  &  1.51  &   0.63  &  1.51  &   0.34 \\
QHC19-D$^*$      &  0.30  & -0.51  & -0.35  &  0.27  &  -0.61  &  0.27  &  -0.90 \\
QMC2         &  0.92  &  0.11  &  0.27  &  0.89  &   0.01  &  0.89  &  -0.28 \\
QMC3         &  0.63  & -0.18  & -0.02  &  0.60  &  -0.28  &  0.60  &  -0.57 \\
QMC4         &  0.56  & -0.26  & -0.09  &  0.52  &  -0.36  &  0.53  &  -0.65 \\
SLY4         &  1.09  &  0.28  &  0.44  &  1.06  &   0.18  &  1.06  &  -0.11 \\
TW           &  0.90  &  0.09  &  0.25  &  0.87  &  -0.01  &  0.87  &  -0.30 \\
\hline
AP4           &  0.68  &  -0.13  &   0.03  &  0.64  &  -0.23  &  0.65  &  -0.52 \\
BSK20         &  0.61  &  -0.20  &  -0.04  &  0.57  &  -0.31  &  0.58  &  -0.59 \\
GMSR-F0       &  0.90  &   0.09  &   0.25  &  0.87  &  -0.01  &  0.87  &  -0.30 \\
GMSR-H1       &  0.32  &  -0.49  &  -0.33  &  0.29  &  -0.59  &  0.29  &  -0.88 \\
GMSR-H2       &  0.29  &  -0.53  &  -0.36  &  0.25  &  -0.63  &  0.26  &  -0.92 \\
GMSR-H4       &  0.32  &  -0.49  &  -0.33  &  0.28  &  -0.59  &  0.29  &  -0.88 \\
GMSR-SLY5     &  0.90  &   0.09  &   0.25  &  0.86  &  -0.02  &  0.87  &  -0.31 \\
SLY230A       &  0.74  &  -0.07  &   0.09  &  0.71  &  -0.17  &  0.71  &  -0.46 \\
SLY2          &  1.05  &   0.24  &   0.40  &  1.02  &   0.14  &  1.02  &  -0.15 \\
SLY           &  1.07  &   0.26  &   0.42  &  1.04  &   0.16  &  1.04  &  -0.13 \\
WFF1          &  2.93  &   2.12  &   2.28  &  2.90  &   2.02  &  2.90  &   1.73 \\
WFF2          &  0.99  &   0.18  &   0.34  &  0.96  &   0.08  &  0.96  &  -0.21 \\
\hline
NL-RMF1      &  0.87  &  0.06  &  0.22  &  0.84  &  -0.04  &  0.84  &  -0.33 \\
NL-RMF2      &  0.65  & -0.16  &  0.00  &  0.62  &  -0.26  &  0.62  &  -0.55 \\
NL-RMF(soft) &  1.22  &  0.40  &  0.56  &  1.18  &   0.30  &  1.19  &   0.01 \\
\hline
$\langle \log_{10}{(B^{SQS}_{NS})}\rangle$ & 0.91 &
 0.10 &
 0.26 &
 0.88 &
 -0.00 &
 0.88 &
 -0.29
\enddata
\end{deluxetable*}


\begin{deluxetable*}{r|ccccccc}
\digitalasset
\tablewidth{0pt}
\tablecaption{Same as Table.~\ref{tab:old_nicer_only} but using posteriors of all NICER measurements (J0740, J0030, J0437, J0614) only. \label{tab:full_nicer_only}. $^*$QHC19-D is a hybrid star EoS}
\tablehead{
\colhead{Model} & \colhead{SQS1} & \colhead{SQS2} & \colhead{SQS3} & \colhead{SQS4} & \colhead{SQS5} & \colhead{SQS6} & \colhead{SQS7}
}
\startdata
APR          &  0.73  &   0.18  &   0.28  &  0.60  &  0.13  &  0.60  &  -1.07 \\
BL           &  1.42  &   0.86  &   0.96  &  1.29  &  0.81  &  1.29  &  -0.38 \\
BSk24        &  1.53  &   0.97  &   1.08  &  1.40  &  0.92  &  1.40  &  -0.27 \\
FSU2R        &  2.56  &   2.00  &   2.10  &  2.43  &  1.95  &  2.43  &   0.75 \\
PCSB2        &  2.45  &   1.89  &   1.99  &  2.32  &  1.84  &  2.32  &   0.65 \\
QHC19-D$^*$      &  0.38  &  -0.18  &  -0.08  &  0.25  & -0.23  &  0.25  &  -1.43 \\
QMC2         &  1.14  &   0.59  &   0.69  &  1.01  &  0.54  &  1.01  &  -0.66 \\
QMC3         &  1.04  &   0.48  &   0.58  &  0.91  &  0.43  &  0.91  &  -0.76 \\
QMC4         &  1.08  &   0.52  &   0.63  &  0.95  &  0.47  &  0.95  &  -0.72 \\
SLY4         &  1.13  &   0.58  &   0.68  &  1.00  &  0.53  &  1.00  &  -0.67 \\
TW           &  1.39  &   0.83  &   0.94  &  1.26  &  0.78  &  1.26  &  -0.41 \\
\hline
AP4           &  0.65  &   0.10  &   0.20  &  0.52  &   0.05  &   0.52  & -1.15 \\
BSK20         &  0.73  &   0.17  &   0.27  &  0.60  &   0.12  &   0.60  & -1.07 \\
GMSR-F0       &  0.95  &   0.39  &   0.49  &  0.82  &   0.34  &   0.82  & -0.85 \\
GMSR-H1       &  0.35  &  -0.21  &  -0.10  &  0.22  &  -0.26  &   0.22  & -1.45 \\
GMSR-H2       &  0.42  &  -0.14  &  -0.04  &  0.29  &  -0.19  &   0.29  & -1.39 \\
GMSR-H4       &  0.54  &  -0.02  &   0.09  &  0.41  &  -0.07  &   0.41  & -1.26 \\
GMSR-SLY5     &  1.00  &   0.45  &   0.55  &  0.88  &   0.40  &   0.87  & -0.80 \\
SLY230A       &  0.88  &   0.32  &   0.42  &  0.75  &   0.27  &   0.75  & -0.93 \\
SLY2          &  1.14  &   0.58  &   0.68  &  1.01  &   0.53  &   1.01  & -0.67 \\
SLY           &  1.16  &   0.60  &   0.70  &  1.03  &   0.55  &   1.03  & -0.64 \\
WFF1          &  2.57  &   2.02  &   2.12  &  2.44  &   1.97  &   2.44  &  0.77 \\
WFF2          &  0.84  &   0.29  &   0.39  &  0.71  &   0.24  &   0.71  & -0.96 \\
\hline
NL-RMF1      &  1.69  &   1.13  &   1.23  &  1.56  &  1.08  &  1.56  &  -0.12 \\
NL-RMF2      &  1.06  &   0.50  &   0.60  &  0.93  &  0.45  &  0.93  &  -0.75 \\
NL-RMF(soft) &  1.23  &   0.67  &   0.78  &  1.10  &  0.62  &  1.10  &  -0.57 \\
\hline
$\langle \log_{10}{(B^{SQS}_{NS})}\rangle$ & 1.16 & 0.60 & 0.70 & 1.03 & 0.55 & 1.03 & -0.65\\
\enddata
\end{deluxetable*}


\begin{deluxetable*}{r|ccccccc}
\digitalasset
\tablewidth{0pt}
\tablecaption{Same as Table.~\ref{tab:old_nicer_only} but using posteriors of all NICER measurements (J0740, J0030, J0437, J0614) as well as the HESS observation. $^*$QHC19-D is a hybrid star EoS\label{tab:full_nicer_hess}}
\tablehead{
\colhead{Model} & \colhead{SQS1} & \colhead{SQS2} & \colhead{SQS3} & \colhead{SQS4} & \colhead{SQS5} & \colhead{SQS6} & \colhead{SQS7}
}
\startdata
APR          &  0.87  &   0.23  &  0.39  &  0.77  &   0.12  &  0.70  &  -0.80 \\
BL           &  2.77  &   2.13  &  2.29  &  2.66  &   2.01  &  2.60  &   1.10 \\
BSk24        &  2.64  &   2.00  &  2.16  &  2.53  &   1.89  &  2.47  &   0.97 \\
FSU2R        &  4.18  &   3.54  &  3.70  &  4.08  &   3.43  &  4.01  &   2.51 \\
PCSB2        &  4.25  &   3.61  &  3.77  &  4.14  &   3.50  &  4.08  &   2.58 \\
QHC19-D$^*$      &  0.51  &  -0.12  &  0.04  &  0.41  &  -0.24  &  0.35  &  -1.15 \\
QMC2         &  1.84  &   1.20  &  1.36  &  1.73  &   1.09  &  1.67  &  0.17 \\
QMC3         &  2.07  &   1.43  &  1.59  &  1.96  &   1.32  &  1.90  &   0.40 \\
QMC4         &  1.78  &   1.14  &  1.30  &  1.67  &   1.03  &  1.61  &   0.11 \\
SLY4         &  1.51  &   0.87  &  1.03  &  1.40  &   0.76  &  1.34  &  -0.16 \\
TW           &  2.76  &   2.12  &  2.29  &  2.66  &   2.01  &  2.60  &   1.09 \\
\hline
AP4           &  0.82  &   0.19  &   0.35  &  0.72  &   0.07  &   0.66  & -0.84 \\
BSK20         &  1.04  &   0.40  &   0.56  &  0.93  &   0.29  &   0.87  & -0.63 \\
GMSR-F0       &  1.28  &   0.64  &   0.80  &  1.17  &   0.53  &   1.11  & -0.39 \\
GMSR-H1       &  0.43  &  -0.21  &  -0.05  &  0.33  &  -0.32  &   0.26  & -1.24 \\
GMSR-H2       &  0.57  &  -0.07  &   0.09  &  0.46  &  -0.19  &   0.40  & -1.10 \\
GMSR-H4       &  0.79  &   0.15  &   0.31  &  0.68  &   0.04  &   0.62  & -0.88 \\
GMSR-SLY5     &  1.58  &   0.94  &   1.10  &  1.48  &   0.83  &   1.42  & -0.09 \\
SLY230A       &  1.31  &   0.67  &   0.83  &  1.20  &   0.55  &   1.14  & -0.36 \\
SLY2          &  1.67  &   1.04  &   1.20  &  1.57  &   0.92  &   1.51  &  0.01 \\
SLY           &  1.77  &   1.13  &   1.29  &  1.66  &   1.02  &   1.60  &  0.10 \\
WFF1          &  2.38  &   1.74  &   1.90  &  2.28  &   1.63  &   2.21  &  0.71 \\
WFF2          &  0.87  &   0.24  &   0.40  &  0.77  &   0.12  &   0.71  & -0.79 \\
\hline
NL-RMF1      &  2.84  &   2.20  &  2.36  &  2.73  &   2.08  &  2.67  &   1.17 \\
NL-RMF2      &  1.96  &   1.32  &  1.48  &  1.85  &   1.20  &  1.79  &   0.29 \\
NL-RMF(soft) &  1.74  &   1.10  &  1.26  &  1.64  &   0.99  &  1.57  &   0.07 \\
\hline
$\langle \log_{10}{(B^{SQS}_{NS})}\rangle$ & 1.78 &
 1.14 &
 1.30 &
 1.67 &
 1.03 &
 1.61 &
 0.11
\enddata
\end{deluxetable*}


\bibliography{refs}



\end{document}